\documentstyle[12pt]{article}

\begin{document}
\setlength{\baselineskip}{5mm}

\noindent{\large\bf Effective Dual Higgs Model with a Dipole-Type
Field

}\vspace{4mm}

\noindent{G.A. Kozlov

}\vspace{1mm}

\noindent{\small

 Bogoliubov Laboratory of Theoretical Physics,
 Joint Institute for Nuclear Research,
 141980 Dubna, Moscow Region, Russia

}\vspace{4mm}



{\small The dual Higgs model is reformulated in terms of two-point Wightman functions
 with the equations of motion involving  higher derivatives.
In the system of the test color charges an analytic expression for the string
 tension is derived.\\


{\bf 1}. The nature of confining forces in gauge theories
attracts a big attention during last years.
 We consider the model (in four-dimensional
space-time (4d)) based on the dual description of a
long-distance Yang-Mills (LDY-M) theory which can provide a
quark confinement in a system of static test charges.
This talk follows
 the idea that the vacuum of quantum Yang-Mills
(Y-M) theory is realized by a condensate of monopole-antimonopole
pairs [1-4].
 Since there are no monopoles as classical
solutions with finite energy in a pure Y-M theory, it has been
suggested by 't Hooft [5] going into the Abelian projection
where the gauge group SU(2) is broken by a suitable gauge
condition to its Abelian subgroup U(1).
Now there is the well-known statement that the interplay between a quark and
antiquark
is analagous to the interaction between a monopole and an antimonopole
 in a superconductor.

 It is known that the topology of the Y-M SU(N) manifold and that of its
Abelian subgroup $[U(1)]^{N-1}$ are different, and
new topological objects can appear in case of
introducing the local gauge transformation of some gauge function
in our model, eg., the field strength tensor for the gauge field
$A_{\mu}(x)$ in quantum chromodynamics (QCD) with
$D_{\mu}(x)=\partial_{\mu}+i\,e\,A_{\mu}(x)$

$$F_{\mu\nu}(x)=\frac{1}{i\,e}\left ([\,D_{\mu}(x), D_{\nu}(x)]- [\,\partial_{\mu},
\partial_{\nu}]\right )\ ,
 $$
which transforms with the gauge function
$\Omega(x)$ as
\begin{eqnarray}
\label{e1}
F_{\mu\nu}(x)\rightarrow
F^{\Omega}_{\mu\nu}(x)=\Omega (x)\,F_{\mu\nu}(x)\,\Omega^{-1} (x)=
\partial_{\mu}\,A^{\Omega}
_{\nu}(x)-\partial_{\nu}\, A^{\Omega}
_{\mu}(x)+ \cr
+ie\,[A^{\Omega}_{\mu}(x),A^{\Omega}_{\nu}(x)]-\frac{1}{i\,e}\,\Omega (x)\,
[\,\partial_{\mu},\partial_{\nu}]\Omega^{-1}(x)\ .
\end{eqnarray}
 The last term in (\ref{e1}) reflects the singular
character of the above-mentioned gauge transformation.
One can identify the Abelian projection by the replacement
$$F^{\Omega}_{\mu\nu}(x)\rightarrow
F^{\alpha}_{\mu\nu}(x)=\partial_{\mu}\,A^{\alpha}
_{\nu}(x)-\partial_{\nu}\, A^{\alpha}
_{\mu}(x)-\frac{1}{i\,e}\,\Omega (x)\,
[\,\partial_{\mu},\partial_{\nu}]\Omega^{-1} (x)\ ,$$
where $A^{\Omega}_{\mu}\rightarrow A^{\alpha}_{\mu}$,
while the label $\alpha$ reflects the Abelian world. This leads to
the Dirac string and magnetic current
$$J^{mon}_{\mu}(x)=-\frac{1}{2\,i\,e}\,\epsilon_{\mu\nu\rho\sigma}\,\partial_{\nu}\,
\Omega(x)[\partial_{\rho},\partial_{\sigma}]\,\Omega^{-1}(x) $$
in the Abelian gauge sector.

Formally, a gauge group element, which transforms a generic SU(3)
connection onto the gauge fixing surface in the space of
connections, is not regular everywhere in spacetime. The
projected (or transformed) connections contain topological
singularities (or defects). Such a singularity
 may form the worldline(s) of magnetic monopoles.
Hence, this singularity leads to the monopole current
$J_{\mu}^{mon}$. This is a natural way of the transformation
from the Y-M theory to a model dealing with Abelian fields.

Analytical models of the dual QCD with
monopoles were intensively investigated [6-8]. The monopole
confinement mechanism is confirmed by many lattice calculations
(see reviews [9] and references therein).
 We study the Lagrangian model
where the fundamental variables are an octet of dual potentials
coupled minimally to three octets of monopole (Higgs) fields.
The dual gauge model is studied at the lowest order of the
perturbative series using the canonical quantization. The basic
manifestation of the model is that it generates the equations of
motion where one of them for the scalar (Higgs) field
is the equation involving higher derivatives.
Our aim is to apply the previous results [10-16] of Wightman
functions in 4d models with equations of motion (for the scalar
fields) involving the dipole-type structure at least, to the
model with monopoles.
The monopole fields
obeying such an equation are classified by their two-point
Wightman functions (TPWF).
In the scheme presented in this talk, the flux distribution in
the tubes formed between two heavy color charges is understood
via the following statement: the Abelian monopoles are excluded
from the string region while the Abelian electric flux is
squeezed into the string region.
 In our model, we use the
dual gauge field $\hat{C}_{\mu}^{a}(x)$ and the monopole field
$\hat{B}_{i}^{a}(x)$ ($i=1,..., N_{c}(N_{c}-1)/2$ and $a$=1,...,8
 is a color index) which
are relevant modes for infrared behaviour. The local coupling of
the $\hat{B}_{i}^{a}$-field to the $\hat{C}_{\mu}^{a}$-field provides
the mass of the dual field and, hence, a dual Meissner effect.
 The commutation relations, TPWF and
Green's functions as well-defined distributions in the space
$S(\Re^{d})$ of complex Schwartz test functions on $\Re^{d}$,
will be defined in the text.

{\bf 2}. Let us consider
the Lagrangian density (LD) $L$ of the $U(1)\times U(1)$ dual Higgs
model corresponding to the LDY-M theory [8]
\begin{eqnarray}
\label{e2}
L=2\,Tr\left [
-\frac{1}{4}\hat{F}^{\mu\nu}\hat{F}_{\mu\nu}+\frac{1}{2}\left
(D_{\mu}\hat{B}_{i}\right )^2\right ] - W\left
(\hat{B}_{i}\right )\ ,
\end{eqnarray}
where
$$ \hat{F}_{\mu\nu}=\partial_{\mu}\hat{C}_{\nu}-\partial_{\nu}\hat{C}_{\mu}-
ig\lbrack\hat{C}_{\mu},\hat{C}_{\nu}\rbrack\ ,$$
$$D_{\mu}\hat{B}_{i}=\partial_{\mu}\hat{B}_{i}-ig\,\lbrack\hat{C}_{\mu},\hat{B}_
{i}\rbrack\ ,$$
$\hat{C}_{\mu}$ and $\hat{B}_{i}$ are the SU(3) matrices, g is
the gauge coupling constant;
$\hat{C}_{\mu}=\lambda^{8}\,C_{\mu}$
($\lambda^{a}$ is the generator of SU(3)).
$ \hat{F}_{\mu\nu}$ becomes the Abelian tensor with $C_{\mu}$
being the dual to an ordinary vector potential.
 The Higgs fields
develop their vacuum expectation values (v.e.v.)
$\hat{B}_{{0}_{i}}$ and the Higgs potential $W(\hat{B}_{i})$
has a minimum at $\hat{B}_{{0}_{i}}$.
The v.e.v.
$\hat{B}_{{0}_{i}}$ produce a color monopole generating current
confining the electric color flux.
The interaction between a quark $Q$ and antiquark $\bar{Q}$ is
provided by $\hat{C}_{\mu}$ via a Dirac string tensor [17] $\hat{\tilde{G}}_{\mu\nu}$
having the same color structure as $\hat{C}_{\mu}$, namely
$\hat{\tilde{G}}_{\mu\nu}=\lambda^{8}\,\tilde{G}_{\mu\nu}$.
 A bound state of $Q$ and $\bar{Q}$ occurs with replacing $
\hat{F}_{\mu\nu}\rightarrow\lambda^{8}\,G_{\mu\nu}$, where
$G_{\mu\nu}=\partial_{\mu}C_{\nu}-\partial_{\nu}C_{\mu}+\tilde{G}_{\mu\nu}$.
Following [8], we choose the color structure for
$\hat{B}_{i}$-fields
$$\hat{B}_{1}=\lambda_{7}\,B_{1}-\lambda_{6}\,\bar B_{1}\, ,$$
$$\hat{B}_{2}=-\lambda_{5}\,B_{2}+\lambda_{4}\,\bar B_{2}\, ,$$
$$\hat{B}_{3}=\lambda_{2}\,B_{3}-\lambda_{1}\,\bar B_{3}\,  .$$
The effective potential looks like (see also [8])
$$ W(B,\bar B, B_{3})=\frac{2}{3}\lambda\{ 11 [ 2 (
B^2 + \bar {B}^2 - B_{0}^2 )^2 + ( B_{3}^2 -
B_{0}^2 )^2 ] $$
$$ +7 [ 2 ( B^2+
\bar{B}^2 ) + B_{3}^2 - 3 B_{0}^2 ]^2\}\ , $$
 where $\lambda$ is dimensionless.
The LD (~\ref{e2}~) becomes
\begin{eqnarray}
\label{e3}
L(\tilde{G}_{\mu\nu})=-\frac{1}{3}G_{\mu\nu}^2+4{\vert
(\partial_{\mu}-igC_{\mu} )
\phi\vert}^2+2(\partial_{\mu}\phi_{3})^2-W(\phi,\phi_{3})\ ,
\end{eqnarray}
where $ \phi\equiv\phi_{1}=\phi_{2}=B-i\bar{B}, \phi_{3}=B_{3};$
 The
generating current of (\ref{e3}) is nothing but the monopole
current confining the electric color flux
$ J^{mon}_{\mu}=(2/3)\,\partial^{\nu}\,G_{\mu\nu}(x)$.
The formal consequence of the $J^{mon}_{\mu}$ conservation,
$\partial^{\mu}J^{mon}_{\mu}=0$, means that monopole currents
form closed loops.

Since $\phi_{1}$ and  $\phi_{2}$ couple to $C_{\mu}$ in the same
way [8], we choose $ B(x)=b(x)\, +\,B_{0}, \bar {B}(x)=\bar{b}(x),
B_{3}(x)=b_{3}(x)\, +\,B_{0}$ with ${\langle B(x)\rangle
}_{0}\not= 0$, ${\langle\bar{B}(x)\rangle}_{0}\not= 0$,
 ${\langle B_{3}(x)\rangle}_{0}\not= 0$.
In terms of the new fields $b, \bar{b}, b_{3}$ the LD (\ref{e3}) is divided into two
parts
$L=L_{1}\,+\,L_{2}$ where $L_{1}$ in the lowest order of $g$ and $\lambda$ and with
the minimal weak interaction looks like
$$L_{1}=-\frac{1}{3}G^{2}_{\mu\nu}\,+\,4\left
[(\partial_{\mu}b)^2+(\partial_{\mu}\bar{b})^2+\frac{1}{2}(\partial_{\mu}b_{3})^2\right
]$$
$$+\,m^2\,C^2_{\mu}-\frac{4}{3}\mu^2
(50b^2+18b_{3}^2)+8m\partial_{\mu}\bar{b}\,C_{\mu}\ .$$
Here, $m\equiv gB_{0}$ and $\mu\equiv\sqrt{2\lambda}B_{0}$.
 The equations of motion for the fields $b, \bar{b}, b_{3}$ and $C_{\mu}$ are
$$ (\Delta^2 +\mu_{1}^2)\,b(x)=0\ ; $$
$$\Delta^2\,\bar{b}(x) + m(\partial\cdot C)=0\ ;$$
$$ (\Delta^2 +\mu_{2}^2)\,b_{3}(x)=0\ ;$$
\begin{eqnarray}
\label{e6}
 (\Delta^2 +m^2_{1})\,C_{\mu}(x) -\partial_{\mu} (\partial\cdot C) +12\,m\,\partial
_{\mu}\bar{b}(x)-\partial^{\nu}\tilde{G}_{\mu\nu}(x)=0\ ,
\end{eqnarray}
where $\mu_{1}^2=(50/3)\,\mu^2, \mu_{2}^2=12\,\mu^2,
m_{1}^2=3\,m^2 $.
 The formal solution of equation (\ref{e6}) looks like
$$ C_{\mu}(x)=\alpha\,\partial^\nu\tilde{G}_{\mu\nu}(x) -
\beta\,\partial_{\mu}\bar{b}(x)\ , $$
with $\alpha\equiv (3\,m^2)^{-1}$, $ \beta\equiv 4/m$.
We obtain that the dual gauge field is defined via the divergence
of the Dirac string tensor $\tilde{G}_{\mu\nu}(x)$ shifted by the divergence
of the scalar field $\bar{b}(x)$.  For large enough $\vec{x}$,
 the monopole field is going to its v.e.v. while
 $C_{\mu}(\vec{x}\rightarrow\infty)\rightarrow 0$ and
$ J_{\mu}^{mon}(\vec{x}\rightarrow\infty)\rightarrow 8\,m^2\,C_{\mu}$.
It implies that in the $d=2h$ dimensions the $\bar{b}(x)$-field
obeys the equation
\begin{eqnarray}
\label{e7}
\Delta^{2h}\,\bar{b}(x)\simeq 0\ ,\,\,\,\,\,\,\, h=2,3,...\ ,
\end{eqnarray}
 but $\Delta^2\,\bar{b}(x)\not=0$.
Here, the solutions of equation (\ref{e7}) obey locality,
Poincare covariance and spectral conditions, and look like the
dipole "ghosts" at h=2.

We define TPWF $W_{h}(x)$ in the d=2\,h-dimensions
$W_{h}(x)={\langle\,\bar{b}(x)\,\bar{b}(0)\,\rangle}_{0}$
as the distribution in the Schwartz space $S^{\prime}(\Re^{2h})$
of temperate distributions on $\Re^{2h}$ which obeys the
equation
\begin{eqnarray}
\label{e9}
\Delta^{2\,h}\,W_{h}(x)=0 \ .
\end{eqnarray}
The general solution of (\ref{e9}) should be Lorentz
invariant and is given in the form [10] at h=2
\begin{eqnarray}
\label{e10}
W_{2}(x)=a_{1}\,\ln\frac{l^2}{-x_{\mu}^2+i\,\epsilon\,x^{0}} +
\frac{a_{2}}{x_{\mu}^2-i\,\epsilon\,x^{0}} + a_{3}\ ,
\end{eqnarray}
where $a_{i}~ (i=1,2,3)$ are the coefficients , $l$ is an arbitrary
length scale. The coefficients $a_{1}$  and $a_{2}$ in (\ref{e10}) can be
fixed using the canonical commutation relations (CCR)
$ \left [C_{\mu}(x),\pi_{C_{\nu}}(0)\right
]_{\vert_{x^{0}=0}}=i\,g_{\mu\nu}\,\delta^{3}(\vec{x})$ and
$\left [\,\bar{b}(x),\pi_{\bar{b}}(0)\right
]_{\vert_{x^{0}=0}}=i\,\delta^{3}(\vec{x})$, respectively, with
$\pi_{C_{\mu}}(x)=-\frac{4}{3}\,G_{0\mu}(x)$ and
$\pi_{\bar{b}}(x)=8\,\left
[\partial^{0}\,\bar{b}(x)+m\,C^{0}(x)\right ]$.
The standard commutator for the scalar field $\bar{b}$ is
$$\left [\,\bar{b}(x),\bar{b}(0)\,\right ]=(2\,\pi)^2\,i\,\left [
4\,a_{1}\,E_{2}(x)+a_{2}\,D_{2}(x)\right ]\ ,$$
where
$E_{2}(x)=(8\,\pi)^{-1}\,sgn(x^0)\,\theta(x^2)$ and
$ D_{2}(x)=(2\,\pi)^{-1}\,sgn(x^0)\,\delta(x^2)$ are taken into
account. The direct calculation leads to
$ a_{1}=(m^2/48\,\pi^2)$ and $a_{2}=-(1/24\,\pi^2)$.

The propagator of the $\bar{b}$-field in $S^{\prime}(\Re_{4})$ is
\begin{eqnarray}
\label{e12}
\hat{\tau}_{2}(p)=weak\lim_{\tilde{\kappa}^{2}<< 1}\,
\frac{i}{3\,(2\,\pi)^4}\,\left\{m^2\left
[\frac{1}{(p^2-\kappa^2+i\,\epsilon)^2}+i\,\pi^2\,\ln\frac{\kappa^2}{\tilde{\mu}^2}\,
\delta_{4}(p)\right ] \right. \cr
\left.
-\frac{1}{2}\,\frac{1}{p^2-\kappa^{2}+i\,\epsilon}\right\}.
\end{eqnarray}
Here, $\kappa$ is a parameter of representation and not an
analogue of the infrared mass,
$\tilde{\kappa}^{2}\equiv\kappa^2/p^2$.

To define the commutation relation $[\,C_{\mu}(x),C_{\nu}(y)\,]$, let us
consider the canonical conjugate pair $\{C_{\mu},\pi_{C_{\nu}}\}$
\begin{eqnarray}
\label{e13}
\left
[\frac{4}{3}C_{\mu}(x),\partial_{\nu}C_{0}(0)-\partial_{0}C_{\nu}(0)-g_{0\nu}
(\partial\cdot C(0))+\Delta_{0\nu}(0)
\right
]_{\vert_{x^{0}=0}}=ig_{\mu\nu}\delta^{3}(\vec{x}),
\end{eqnarray}
where $\Delta_{\mu\nu}(x)=g_{\mu\nu}\,(\partial\cdot
C(x))-\tilde{G}_{\mu\nu}(x)$ tends to zero as $x\rightarrow
0$ and the Dirac string tensor $\tilde{G}_{\mu\nu}(x)$ obeys the
equation $ \left [\Delta^2 + (3m)^2\right
]\,\tilde{G}_{\mu\nu}(x)=0$. Obviously, the following form of
 the free $C_{\mu}$-field commutator:
\begin{eqnarray}
\label{e14}
\left
[\,C_{\mu}(x),C_{\nu}(0)
\right
]=ig_{\mu\nu}\left [\,\xi\,m_{1}^2\,E_{2}(x)+c\,D_{2}(x)\right ]\ ,
\end{eqnarray}
ensures the CCR (\ref{e13}) at large $x_{\mu}^2$ with both
$\xi$ and $c$ (in (\ref{e14})) being real arbitrary numbers but
$\xi=\frac{3}{4}-4\,c$.

The free dual gauge field propagator in $S^{\prime}(\Re_{4})$ in
any local covariant gauge is given by
\begin{eqnarray}
\label{e15}
\hat{\tau}_{\mu\nu}(p)=\int\,d^{4}x\,\exp(i\,p\,x)\,\tau_{\mu\nu}(x)
\cr
=i\,\left [g_{\mu\nu}-\left (1-\frac{1}{\zeta}\right
)\,\frac{p_{\mu}\,p_{\nu}}{p^2+i\,\epsilon}\right ]\cdot\left
[\xi\,m_{1}^2\,\hat{t}_{1}(p)+c\,\hat{t}_{2}(p)\right ]\ ,
\end{eqnarray}
where
$$ \tau_{\mu\nu}(x)=\frac{i\,g_{\mu\nu}}{(4\,\pi)^2}\,\left
[\,\xi\,m_{1}^2\,\ln(-\tilde{\mu}^2\,x_{\mu}^2+i\,\epsilon
)+\frac{c}{x_{\mu}^2+i\,\epsilon}\right ]\ ; $$

$$ \hat{t}_{1}(p)=weak\,\lim_{\tilde{\kappa}^2 <<1}\,\left
[\frac{1}{(p^2-\kappa^2+i\,\epsilon)^2} +
i\,\pi^2\,\ln\left(\frac{\kappa{^2}}{\tilde{\mu}^2}\right
)\,\delta_{4}(p)\right ]\ ; $$

$$ \hat{t}_{2}(p)=weak\,\lim_{\tilde{\kappa}^2 <<1}\, \frac{1}{2}\,
\frac{1}{(p^2-\kappa{^2}+i\,\epsilon)} \ . $$
The gauge parameter $\zeta$ in (\ref{e15}) is a real number.
 The following requirement
$ (\Delta^{2})^{2}\,\tilde{\tau}_{\mu\nu}(x)=i\,\delta_{4}(x)$
on Green's function $\tau_{\mu\nu}(x)$
leads to that a constant $c$ has to be equal to zero
and $\tilde{\tau}_{\mu\nu}(x)=\tau_{\mu\nu}(x)/(\xi\,m_{1}^2)$.

{\bf 3}. As for an approximate topological solution
for this dual model, we fix the equations of motion
\begin{eqnarray}
\label{e16}
 \partial^{\nu}\,G_{\mu\nu}=6\,i\,g[\phi^{\ast}
(\partial_{\mu}-i\,g\,C_{\mu})\,\phi\,-\,\phi (\partial_{\mu} +
i\,g\,C_{\mu})\,\phi^{\ast} ]\ ,
\end{eqnarray}
\begin{eqnarray}
\label{e17}
(\partial_{\mu}-i\,g\,C_{\mu})^2\,\phi=\frac{2}{3}\,\lambda\,(32\,B_{0}^2-25\,{\vert
\,\phi\,\vert}^2-7\,\phi_{3}^2\,)\,\phi\ ,
\end{eqnarray}
where $\phi(x)$ is decomposed like
$\phi(x)=\frac{1}{\sqrt{2}}\,\exp(i\,f(x))\,[\,\chi(x)+B_{0}\,]$
 using the new scalar variables $\chi(x)$ and $f(x)$.
The equation of motion (\ref{e16}) transforms into the following
one:
$$\partial^{\nu}\,G_{\mu\nu}=6\,g\,(\chi+B_{0})^2\,(g\,C_{\mu}-\partial_{\mu}\,f)\
,$$
that means that the $\bar{b}(x)$-field is nothing but a
mathematical realization of the "massive" phase $B_{0}\cdot f(x)$
at large enough $\vec{x}$,
i.e., $\bar{b}(x)\simeq (B_{0}/2)\,S(x)\,f(x)$
and $S(x)\equiv (1+\chi(x)/B_{0})^2$.
Integrating out $G_{\mu\nu}$ over the 2d surface element $\sigma^{\mu\nu}$
in the flux $\Pi=\int\,G_{\mu\nu}(x)\,d\,\sigma^{\mu\nu}$, we
conclude that the phase $f(x)$ is varied by
$2\,\pi\,n$ for any integer number $n$ associated with the
topological charge [18] inside the flux tube.
Using the cylindrical symmetry we arrive at the field equation $
(\vec{C}\rightarrow (\tilde{C}(r)/r)\vec{e_{\theta}})$
$$\frac{d^2\,\tilde{C}(r)}{d\,r^2}-\frac{1}{r}\,\frac{d\,\tilde{C}(r)}{d\,r}-
3\,m^2\,[\,3+2\,S(r)]\,\tilde{C}(r)+6\,n\,m\,B_{0}\,S(r)=0\ $$
with the asymptotic transverse behaviour of its solution
$$
\tilde{C}(r)\simeq\frac{4\,n}{7\,g}-\sqrt{\frac{\pi\,m\,r}{2\,k}}\,e^{-k\,m\,r}\,
\left (\,1+\frac{3}{8\,k\,m\,r}\right )\ ,\,\,\,\,
k\equiv\sqrt{21}\ . $$
The field equation (\ref{e17}) is given by
$$\frac{d^2\,\hat\chi(r)}{d\,r^2}+\frac{1}{r}\,\frac{d\,\hat\chi(r)}{d\,r}-
\left\{\left[\frac{n-g\,\tilde{C}(r)}{r}\right]^{2}+\frac{25}{3}\lambda
\left[2\,B_{0}^2-\hat\chi^{2}(r)\right]\right\}\hat\chi(r)\simeq 0\
, $$
where $\hat\chi(r)=\chi(r)+B_{0}$.
The profile of the color electric field in the flux tube at large
$r$ looks like
$$ E_{z}(r)=\sqrt{\frac{\pi\,m}{2\,k\,r}}\,\left
(k\,m-\frac{1}{2\,r}\right )\,e^{-k\,m\,r}\ . $$

{\bf 4}. Now, our aim is to obtain the confinement potential
in an analytic form for the system of interacting
static test charges of a quark and an antiquark.
According to the distribution (\ref{e12}), the static potential
in $\Re^{3}$ is a rising function with $r=\vert\vec{x}\vert$
[14,15]
$$P_{stat}(r)\sim\frac{1}{2^{2\,h}\,\pi^{3/2}}\,\frac{1}{(h-1)!}\,
 \Gamma (3/2-h)\,r^{2\,h-3}\ ,$$
and the Fourier transformation of the analytic function
$r^{\sigma}$ at $\sigma\neq -d, -d-2,...$ is [15]

$$ F\{r^{\sigma}\}=\left (\frac{4\,\pi}{p^2}\right
)^{(\sigma+d)/2}\,\pi^{-\sigma/2}\,\frac{\Gamma [(\sigma+d)/2]}{\Gamma (-\sigma
/2)}\ . $$
In this talk, we define the static potential like
$P_{stat}(r)=\lim_{T\rightarrow\infty}\,[T^{-1}\cdot A(r)]$ and
the action $A(r)$ is given by the colour source-current
part of LD
$L(p)=-\vec{j}_{\alpha}^{\mu}(-p)\,\hat{\tau}_{\mu\nu}(p)\,\vec{j}_{\alpha}^{\nu}(p)\ $
with the quark current $ \vec{j}_{\alpha}^{\mu}(x)=\vec{Q}_{\alpha}\,g^{\mu
0}\,[\delta_{3}(\vec{x}-\vec{x_{1}})-\delta_{3}(\vec{x}-\vec{x_{2}})]$.
Here, $\vec{Q}_{\alpha}=e\,\vec{\rho}_{\alpha}$ is the Abelian
color-electric charge of a quark while $\vec{\rho}_{\alpha}$ is
the weight vector of the SU(3) algebra: $\rho_{1}=(1/2,\,\sqrt{3}/6)$,
$\rho_{2}=(-1/2,\,\sqrt{3}/6)$, $\rho_{3}=(0,\,-1/\sqrt{3})$
[18]; $\vec{x}_{1}$ and $\vec{x}_{2}$ are the position vectors
of a quark and an antiquark, respectively.

As a consequence of the dual field propagator (\ref{e15}), and
using the following representation in the sense of generalized
functions [11]
$$weak \lim_{{\tilde{\kappa}}^2 <<1}\,\left
[\frac{1}{(p^2-\kappa^2+i\,\epsilon)^2}+i\,\pi^2\,\ln\frac{\kappa^2}{\tilde{\mu}^2}\,
\delta_{4}(p)\right ]= $$
$$=\frac{1}{4}\frac{\partial^2}{\partial\,p^2}\,\frac{1}{-p^2-i\,\epsilon}\,\ln
\frac{-p^2-i\,
\epsilon}{\tilde{\mu}^2}=\frac{1}{2}\,\frac{1}{(p^2+i\,\epsilon)^2}\,\left
(5-3\,\ln\frac{-p^2-i\,
\epsilon}{\tilde{\mu}^2}\right )\ ,$$
we get
\begin{eqnarray}
\label{e22}
P_{stat}(r)=-\frac{\vec{Q}^2}{16\,\pi}\left
[\xi\,m^2\,r(-12.4+6\,\ln\tilde{\mu}\,r
)+O\left(\frac{c}{r}\right )\right] \ .
\end{eqnarray}
Hence, the string tension $a$ in $P_{stat}(r)\sim a\,r$ emerges as
$$a\simeq \frac{3}{4}\,\frac{\pi}{g^2}\,m^2 = \frac{3}{4}\,\alpha_{s}\,m^2\ , $$
where the logarithmic correction in (\ref{e22}) has been
ignored at large distances $r\sim{\tilde\mu}^{-1}$.
 We obtain $a\simeq 0.18~GeV^2$ for the mass of the dual
$C_{\mu}$-field $m=0.8~GeV$ and $\alpha_{s}= \pi/g^{2}=e^2/(4\,\pi)$=0.37
fixed from fitting the heavy quark-antiquark spectrum [19].
The string tension $a_{fixed}=1.0~GeV/fm$ extracted from the
Regge slope of the hadrons [20] then determines the fixed value
of the mass $m_{fixed}=0.85~GeV$.
Doing the formal comparison, let us remind that the string tensions in
papers [21] and [22] are equal to $a=(430~MeV)^{2}$ and
$a=(440~MeV)^{2}$,
respectively.
 We found that for a
sufficiently long string the $\sim r$-behaviour of
the static potential is dominant; for a short string
the singular interaction provided by the second term in (\ref{e22})
becomes important if the average size of the monopole is even
smaller.

{\bf 5}. Finally, some conclusion is in order

a). We have actually derived the analytic expressions of both
the $\bar{b}$-field (\ref{e12}) and the dual gauge boson field
(\ref{e15}) propagators in $S^{\prime}(\Re_{4})$. Our result
should be regarded as the distributions (\ref{e12}) and
(\ref{e15}) in a weak sense.

b). The dual gauge bosons become
massive due to their interaction with scalar field(s). But not
every scalar species becomes massive since the symmetry breaking
pattern is $SU(3)\rightarrow U(1)\times U(1)$ and one scalar
field remains massless.
 We see that the fields
$b(x)$ and $b_{3}(x)$ receive their masses and the $\bar{b}(x)$
field in combination with $\partial^{\nu}\,\tilde{G}_{\mu\nu}(x)$ form
the vector field $C_{\mu}(x)$ obeying the equation of motion for
the massive vector field with the mass $m=g\,B_{0}$.
The solution of the $\bar{b}(x)$-field can be identified as a
 "ghost"-like particle in the substitute manner.

c). The form of the
potential (\ref{e22}) grows linearly with the distance $r$ apart from a
logarithmic correction.

d). The string tension $a$ obtained is quite close to a
phenomenological value.

e). Since no real physics can depend on the choice of the gauge
group (where the Abelian group appears as a subgroup) it seems
to be a new mechanism of confinement [23].\\

\newcommand{\etal}{{\em et al.}}
\setlength{\parindent}{0mm}
\vspace{5mm}
{\bf References}
\begin{list}{}{\setlength{\topsep}{0mm}\setlength{\itemsep}{0mm}%
\setlength{\parsep}{0mm}}
%

\item[1.] Y. Nambu, Phys. Rev. D {\bf 10}, 4262 (1974) .
\item[2.] S. Mandelstam, Phys. Rep. C {\bf 23}, 245 (1976); Phys. Rev. D {\bf 19},
2391 (1979).
\item[3.] A.M. Polyakov, Nucl. Phys. B {\bf 120}, 429 (1977).
\item[4.] G.'t Hooft, Nucl. Phys. B {\bf 138}, 1 (1978).
\item[5.] G.'t Hooft, Nucl. Phys. B [FS3] {\bf 190}, 455 (1981).
\item[6.] T. Suzuki, Prog. Theor. Phys. {\bf 80}, 929 (1988); {\bf 81}, 752
(1989); S. Maedan and T. Suzuki, Prog. Theor. Phys. {\bf 81},
229 (1989).
\item[7.] H. Suganuma, S. Sasaki and H. Toki, Nucl. Phys. B {\bf 435}, 207
(1995).
\item[8.] M. Baker et al., Phys. Rev. D {\bf 54}, 2829 (1996).
\item[9.] T. Suzuki, Nucl. Phys. (Proc. Suppl.) {\bf 30}, 176 (1993);
M.I. Polikarpov, Nucl. Phys. (Proc. Suppl.) {\bf 53}, 134 (1997);
A. Di Giacomo, Prog. Theor. Phys. Suppl. {\bf 131}, 161 (1998);
M.N. Chernodub et al., Prog. Theor. Phys. Suppl. {\bf 131}, 309 (1998);
R.W. Haymaker, Phys. Rep. {\bf 315}, 153 (1999).
\item[10.] R. Ferrari, Il Nuovo Cimento A {\bf 19}, 204 (1974).
\item[11.] D. Zwanziger, Phys. Rev. D {\bf 17}, 457 (1978).
\item[12.] E. D'Emilio and M. Mintchev, "A gauge model with
perturbative confinement in four dimensions", report IFUP, Pisa
(1979).
\item[13.] E. D'Emilio and M. Mintchev, Phys. Lett. B {\bf 89}, 207 (1980).
\item[14.] G.A. Kozlov, S.P. Kuleshov, and R.P. Zaikov, "On linear
quasipotential in the framework of quantum field theory", report
E2-80-352, JINR, Dubna (1980).
\item[15.]  G.A. Kozlov, IL Nuovo Cimento A {\bf 105}, 139 (1992).
\item[16.] G.A. Kozlov, IL Nuovo Cimento A {\bf 107}, 819, 1739 (1994).
\item[17.] P.A.M. Dirac, Phys. Rev. {\bf 74}, 817 (1948).
\item[18.]  Y. Koma, H. Suganuma, and H. Toki, Phys. Rev. D {\bf 60}, 074024
(1999).
\item[19.] M. Baker, J.S. Ball and F. Zachariasen, Phys. Rev. D {\bf 51}, 1968 (1995).
\item[20.] K. Sailer et al., J. Phys. G {\bf 17}, 1005 (1991).
\item[21.] M. Baker, N. Brambilla, H.G. Dosch, and A. Vairo, Phys. Rev. D {\bf 58},
034010 (1998).
\item[22.] M.N. Chernodub, "Classical string solutions in
effective infrared theory of SU(3) gluodynamics", report
ITEP-TH-28/99.
\item[23.] L.D. Faddeev, A.J. Niemi, Phys. Rev. Lett. {\bf 82}, 1624
(1999); Phys. Lett. B {\bf 449}, 214 (1999);
 S.V. Shabanov, Phys. Lett. B {\bf 463}, 263 (1999);
Y.M. Cho, Phys. Rev. D {\bf 62}, 074009 (2000).

%
\end{list}

\end{document}